\documentclass[11pt]{article}
\usepackage[utf8]{inputenc}
\usepackage[T1]{fontenc}
\usepackage[margin=1in]{geometry}
\usepackage{amsmath}
\usepackage{booktabs}
\usepackage{graphicx}
\usepackage{flafter}
\usepackage[section]{placeins}
\usepackage{tikz}
\usetikzlibrary{arrows.meta,positioning,calc,decorations.pathreplacing}
\usepackage{pgfplots}
\pgfplotsset{compat=1.18}
\usepackage[hidelinks]{hyperref}

\newcommand{\gv}{\ensuremath{\mathbf{g}}}
\newcommand{\sg}{\ensuremath{s_{\mathrm{g}}}}
\newcommand{\thB}{\ensuremath{\theta_{\mathrm{B}}}}
\newcommand{\dg}{\ensuremath{^{\circ}}}

\title{GRADAR: orientation-map-driven detection and ranking of grains for targeted
two-beam electron channeling contrast imaging}
\author{Johan Westraadt\thanks{Center for Electron Microscopy and Analysis (CEMAS),
The Ohio State University, Columbus, OH, USA. Email: \texttt{westraadt.1@osu.edu}.
ORCID: 0000-0002-5951-6955.}}
\date{August 2026}

\begin{document}
\maketitle

\begin{abstract}
\noindent Electron channeling contrast imaging (ECCI) resolves individual dislocations in
bulk samples, but only when the imaged grain is held in a two-beam diffraction condition to
well under half a degree. In a polycrystal it must be re-established grain by grain, and
every published workflow we are aware of chooses the grain first and solves the stage for
that single orientation. This paper treats the inverse problem: nominate one reflection
family and a stage envelope, and determine which grains of an EBSD-mapped polycrystal can
reach that family's two-beam condition, a question posed in the literature in 2019 and open
since. At fixed tilt $t$, stage rotation sweeps the incident beam around a cone of
half-angle $t$ in each grain's crystal frame, and a family with Bragg angle
$\theta_{\mathrm{B}}$ is reachable exactly when one of its plane normals lies within
$(t + \theta_{\mathrm{B}})$ of the specimen surface. On a measured 163-grain austenitic
stainless steel map at 20~kV, $\{111\}$ reachability rises from 38.7\% at $7^{\circ}$ tilt
to 77.3\% at $15^{\circ}$; $\{220\}$ and $\{311\}$ already reach 79.8\% and 91.4\% at
$7^{\circ}$. Ranking survivors by predicted darkness alone is wrong:
same-family candidates are equivalent with respect to their own band at exact Bragg
incidence, and the darkest tend to lie closest to rival-band Bragg conditions. GRADAR
therefore ranks \emph{darkest-clean}, gating every candidate on a minimum angular clearance
from all rivals, and returns a dial-ready stage move per selected grain. Because the beam
moves only about $0.12^{\circ}$ per degree of stage rotation at $7^{\circ}$ tilt, an
ordinary rotation stage positions the selected condition to better than $0.1^{\circ}$. The
predicted sweep is checked against the published silicon precession series of the AstroECP
dataset: simulated and measured traces agree at $r = 0.84$, and the measured intensity is
dark at all 16 predicted two-beam edges. Per-grain validation on polycrystals is deferred to
a companion study.
\end{abstract}

\section{Introduction}
\label{sec:intro}

Electron channeling contrast imaging has moved well past its demonstration phase. Recent work
images dislocation networks tomographically \cite{weidner2026}, follows strain localization at
grain boundaries during in-situ fatigue \cite{su2025}, characterizes creep-deformed olivine
\cite{qaiser2026olivine}, threading dislocations in epitaxial GaN \cite{frascaroli2026} and
III--V films on silicon \cite{monge2026}, irradiation-induced loops \cite{shen2024}, and
mechanical Dauphin{\'e} twins in naturally deformed quartz \cite{miyajima2024}. Every one of these polycrystalline applications
repeats the same preparatory step. For each grain of interest, the operator must find a stage
position that excites exactly one strong reflection (a two-beam condition) before any defect
image is worth recording \cite{zaefferer2014}. The hunt is done by eye on a channeling
pattern, or by trial reorientation, and it starts over with every new grain.

The precision stakes of that step are now well quantified. Qaiser et al.\ conclude that
reliable high-resolution ECCI demands beam--crystal alignment well below half a degree, where
conventional controlled-ECCI workflows often exceed $0.5\dg$ \cite{qaiser2026}. Their
tri-crystal experiment makes the consequence concrete: predicting a channeling pattern from an
EBSD-measured misorientation missed by ${\sim}0.74\dg$, enough to reassign the band under the
optic axis from $\{040\}$ to $\{620\}$, an error that would silently invalidate a
$\gv \cdot \mathbf{b}$ analysis built on it \cite{qaiser2026}. A deviation of only $0.5\dg$
visibly changes the contrast of individual threading dislocations \cite{qaiser2026}. A
related limit applies to the crystal itself: a heavily deformed grain carries large
internal lattice rotations, so no single stage position holds all of it in one channeling
condition. L'h\^{o}te et al.\ measured local disorientations of ${\sim}2\dg$ even in
as-received copper \cite{lhote2019}. Any targeting method that assigns one orientation per
grain, the present one included, is limited to the part of the grain that orientation
represents. For polycrystals carrying large deformations the internal spread, not the
prediction accuracy, is therefore the limiting factor; L'h\^{o}te et al.\ met the same
physics from the acquisition side, averaging images over a range of rotations so that every
part of the grain passes through contrast somewhere in the series \cite{lhote2019}.

Software support for this step exists, and all of it shares one structure. TOCA computes the
tilt and rotation that establish a two-beam condition for a named set of lattice planes, in
one operator-chosen grain \cite{gutierrez2013,zaefferer2014}. openECCI ingests a full
orientation map, yet uses it as a clickable picture from which one grain is selected for a
stage solution \cite{xu2024}. Mori et al.\ compute EBSD-derived tilt and rotation on a
dedicated piezo stage, for the grain under study \cite{mori2024}. On the TEM side, ATEX
solves a named reflection into two-beam within holder limits \cite{wang2026}, ALPHABETA and
$\tau$ompas screen reachable conditions against stage limits \cite{cautaerts2019,xie2020},
and crystalAligner treats constrained stage kinematics by global optimization
\cite{niessen2020}, each for a single given crystal. The loop, everywhere, runs over
reflections within one grain.

The inverse loop was named as an open problem by L'h\^{o}te et al.\ in 2019
\cite{lhote2019}. Introducing rotational ECCI on a copper single crystal, they marked the
candidate two-beam rotations by eye on a simulated channeling pattern, and noted that
simulation would be required to guarantee that a suitable two-beam condition exists for
every crystal orientation at a given rotation, a question they deferred beyond the scope of
their article \cite{lhote2019}. It has not been addressed since. The same lineage has since scaled its orientation
mapping to large areas \cite{lafond2020}, extended it to interfaces \cite{langlois2024}, and
rebuilt its indexing engine \cite{lhote2025}: all refinements of the inverse
(orientation-recovery) direction. We are not aware of a published tool, in that lineage or
any other, that answers the forward question: given the map, which grains can reach a
nominated condition?

This paper answers it. The operator declares a reflection family and a stage envelope, and
GRADAR (G-vector Rotational Azimuthal Detection And Ranking) returns the grains that can
reach that family's two-beam condition, ranked by
predicted channeling contrast subject to a computed clearance against rival reflections,
each with the stage rotation and tilt to dial. The criterion that drives it reduces to two
lines of geometry, checkable by hand; its consequences on a real orientation map are
quantified in section~\ref{sec:results}, and the predicted sweep is checked against a
published measured precession series in section~\ref{sec:validation}. The intended
application makes the inversion natural: high-throughput dislocation analysis in known
material systems, where the informative diffraction conditions are established in advance
and the operator's real choice is the diffraction vector, not the grain. With $\gv$
nominated once, the question becomes which grains can serve it and at what stage settings;
answering that map-wide replaces the per-grain hunt with a computed session plan, so that
statistically representative defect populations can be collected under one declared
condition and compared across material states arising from different processing or testing
histories, a step toward automating ECCI itself.

\section{Reachability under stage rotation at fixed tilt}
\label{sec:geometry}

\subsection{The rotation cone and the two-beam crossing}

\begin{figure}[tbp]
\centering
\includegraphics[width=\textwidth]{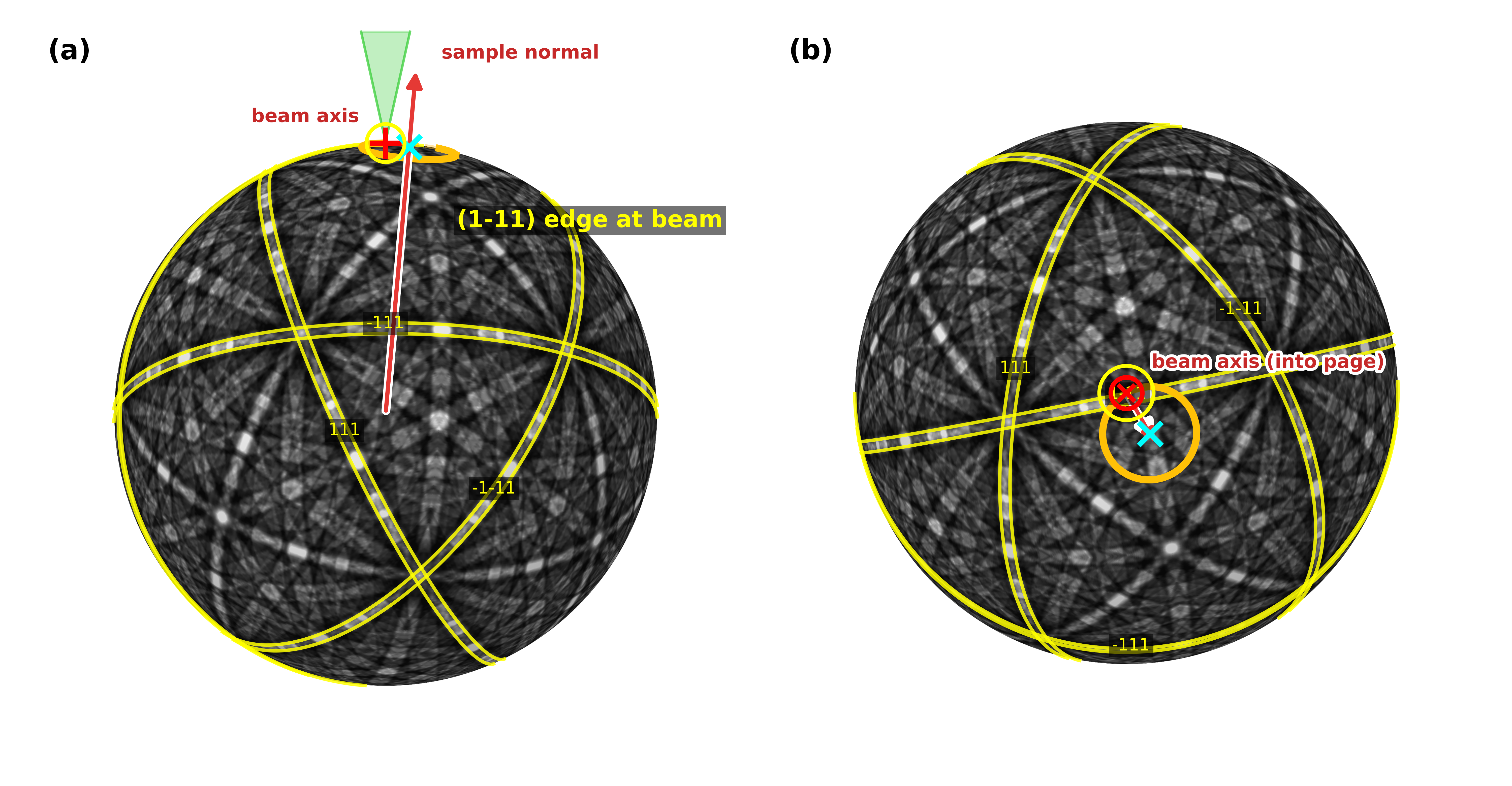}
\caption{Rotational ECCI on the Kikuchi sphere: a dynamically simulated nickel reference
pattern (20~kV) at $10\dg$ stage tilt, frozen the instant a $(1\bar{1}1)$ band edge sweeps
through the beam ($\varphi = 189.5\dg$). (a)~Side view: the incident beam (green cone)
enters at the red cross, the sample normal (red arrow) is tilted $10\dg$ from it, the amber
circle is the beam's trace under stage rotation, and the yellow pairs are the $\{111\}$
band-edge (Kossel) circles, with the $(1\bar{1}1)$ edge highlighted. (b)~The same instant
viewed along the beam axis (circled cross): the pattern precesses while the beam point
stays fixed, and the $(1\bar{1}1)$ edge meets it at the $\sg = 0$ two-beam condition.}
\label{fig:sphere}
\end{figure}

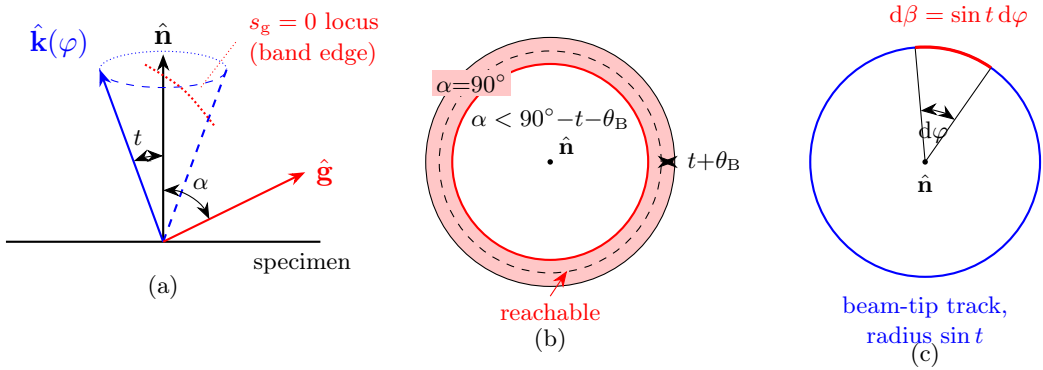
\begin{figure}[tbp]
\centering
\begin{tikzpicture}[scale=0.80, >={Stealth[length=2.2mm]}]
\begin{scope}[xshift=0cm]
  \draw[thick] (-2.6,0) -- (2.6,0);
  \node[below] at (2.3,-0.02) {\footnotesize specimen};
  \draw[->,thick] (0,0) -- (0,3.1) node[above] {$\hat{\mathbf{n}}$};
  \draw[->,thick,blue] (0,0) -- (-1.05,2.85) node[above left] {$\hat{\mathbf{k}}(\varphi)$};
  \draw[thick,blue,dashed] (0,0) -- (1.05,2.85);
  \draw[blue,dashed] (-1.05,2.85) arc[x radius=1.05, y radius=0.3, start angle=180, end angle=360];
  \draw[blue,densely dotted] (-1.05,2.85) arc[x radius=1.05, y radius=0.3, start angle=180, end angle=0];
  \draw[<->] (0,1.42) arc[radius=1.42, start angle=90, end angle=110];
  \node at (-0.42,1.72) {\footnotesize $t$};
  \draw[->,thick,red] (0,0) -- (2.35,1.15) node[right] {$\hat{\mathbf{g}}$};
  \draw[<->] (0,0.85) arc[radius=0.85, start angle=90, end angle=26];
  \node at (0.62,0.95) {\footnotesize $\alpha$};
  \draw[red,densely dotted,thick] (-0.30,2.93) .. controls (0.35,2.55) .. (0.86,1.90);
  \node[red,align=left,anchor=west] at (1.25,3.35) {\footnotesize $\sg=0$ locus\\[-2pt]\footnotesize (band edge)};
  \draw[red,densely dotted] (1.20,3.30) -- (0.62,2.55);
  \node at (0,-0.75) {\footnotesize (a)};
\end{scope}
\begin{scope}[xshift=6.4cm,yshift=1.32cm]
  \draw[thick] (0,0) circle (2.05);
  \fill[red!22,even odd rule] (0,0) circle (2.05) circle (1.62);
  \draw[red,thick] (0,0) circle (1.62);
  \draw[black,dashed] (0,0) circle (1.835);
  \node[fill=red!22, inner sep=1.5pt] at (-1.30,1.30) {\footnotesize $\alpha{=}90\dg$};
  \fill (0,0) circle (0.045) node[above right=-1pt] {\footnotesize $\hat{\mathbf{n}}$};
  \node at (0,0.72) {\footnotesize $\alpha < 90\dg{-}t{-}\thB$};
  \node[red,anchor=north] at (0,-2.18) {\footnotesize reachable};
  \draw[red,->] (0,-2.22) -- (0.28,-1.80);
  \draw[<->] (1.835,0) -- (2.05,0);
  \node[right] at (2.10,0.02) {\footnotesize $t{+}\thB$};
  \node at (0,-2.95) {\footnotesize (b)};
\end{scope}
\begin{scope}[xshift=12.6cm,yshift=1.32cm]
  \draw[thick,blue] (0,0) circle (1.9);
  \fill (0,0) circle (0.045) node[below=1pt] {\footnotesize $\hat{\mathbf{n}}$};
  \node[blue,align=center,anchor=north] at (0,-2.05) {\footnotesize beam-tip track,\\[-2pt]\footnotesize radius $\sin t$};
  \draw (0,0) -- (55:1.9);
  \draw (0,0) -- (95:1.9);
  \draw[<->] (55:0.85) arc[radius=0.85, start angle=55, end angle=95];
  \node at (75:0.52) {\footnotesize $\mathrm{d}\varphi$};
  \draw[very thick,red] (55:1.9) arc[radius=1.9, start angle=55, end angle=95];
  \node[red,align=center,anchor=south] at (75:2.15) {\footnotesize $\mathrm{d}\beta = \sin t\,\mathrm{d}\varphi$};
  \node at (0,-3.20) {\footnotesize (c)};
\end{scope}
\end{tikzpicture}
\caption{The reachability construction. (a)~At fixed tilt $t$, stage rotation $\varphi$
sweeps the incident beam $\hat{\mathbf{k}}$ around a cone of half-angle $t$ about the specimen
normal $\hat{\mathbf{n}}$, drawn in the crystal frame of one grain. The cone crosses the
$\sg = 0$ locus of a reflection $\gv$ (equation~\ref{eq:crossing}) only if the plane normal's
polar angle $\alpha$ satisfies equation~\ref{eq:window}. (b)~On the orientation sphere the
criterion is an annulus of full width $2(t + \thB)$ straddling the dashed centreline
$\alpha = 90\dg$, the locus of plane normals lying exactly in the specimen surface
(stereographic sketch, width exaggerated). (c)~The gearing construction, viewed along
$\hat{\mathbf{n}}$:
the beam tip moves on a circle of radius $\sin t$, so a stage-rotation increment
$\mathrm{d}\varphi$ moves the beam direction by only $\mathrm{d}\beta = \sin t\,
\mathrm{d}\varphi$. All panels are schematic.}
\label{fig:cone}
\end{figure}

Fix the stage tilt at $t$ and rotate the stage through the azimuth $\varphi$. In the specimen
frame the incident beam direction is constant; in the crystal frame of any one grain it is the
specimen that is fixed and the beam that appears to precess. Figure~\ref{fig:sphere} shows
the construction on the Kikuchi sphere of a dynamically simulated nickel reference pattern
at $t = 10\dg$, a still frame from the animated version provided as supplementary
movie~S1 (crystal orientation Euler angles
$81.1\dg$, $27.0\dg$, $314.5\dg$): the beam enters the sphere at a fixed point, and as the
stage rotates about the sample normal the beam direction traces the amber circle, a small
circle of angular radius equal to the tilt, centred on the normal and passing through the
beam entry point; viewed along the beam axis instead (figure~\ref{fig:sphere}b), the same
rotation is a precession of the pattern beneath the stationary beam point. As $\varphi$ runs
through a full turn, the unit beam vector $\hat{\mathbf{k}}(\varphi)$ thus sweeps a cone of
half-angle $t$ about the specimen normal $\hat{\mathbf{n}}$, expressed in that grain's
crystal coordinates and drawn schematically in figure~\ref{fig:cone}a. Every grain of the
map shares the same cone geometry; each sees it rotated by its own orientation.

A reflection $\gv$ with Bragg angle $\thB$ reaches its two-beam condition, $\sg = 0$, when the
beam lies exactly $\thB$ off the $(hkl)$ plane, that is
\begin{equation}
\hat{\mathbf{k}}(\varphi) \cdot \hat{\mathbf{g}} \;=\; \pm \sin\thB ,
\label{eq:crossing}
\end{equation}
with the two signs exciting $+\gv$ and $-\gv$ respectively. These are physically distinct
conditions on the two edges of the same Kikuchi band, and worth keeping distinct because
$\gv \cdot \mathbf{b}$ invisibility analysis depends on which edge is excited
\cite{qaiser2026}. On the Kikuchi sphere the two edges are the Kossel circles at
$90\dg \mp \thB$ from the plane normal, the yellow circle pairs of figure~\ref{fig:sphere}
($\thB = 1.21\dg$ for nickel $\{111\}$ at 20~kV); the frozen instant of that figure is such
a crossing, the $(1\bar{1}1)$ edge coinciding with the beam (supplementary movie~S1 shows
the full sweep). Writing $\alpha$ for the polar
angle of $\hat{\mathbf{g}}$ from the
specimen normal, the left side of equation~\eqref{eq:crossing} ranges over
$[\cos(\alpha + t),\, \cos(\alpha - t)]$ as $\varphi$ sweeps the cone. A solution therefore
exists if and only if
\begin{equation}
90\dg - t - \thB \;\le\; \alpha \;\le\; 90\dg + t + \thB .
\label{eq:window}
\end{equation}
Geometrically, equation~\eqref{eq:window} is an annulus on the orientation sphere: the plane
normal must lie within $(t + \thB)$ of the specimen surface (figure~\ref{fig:cone}b). A family
$\{hkl\}$ is reachable in a given grain when at least one of its symmetric equivalents falls
inside the annulus.

The window is narrow, and that is the entire content of the targeting problem. For a single
plane normal uniformly distributed on the sphere, the probability of landing in the annulus
is $\sin(t + \thB)$, and the fraction of grains with a reachable family condition is bounded
above by the sum of these probabilities over the family's symmetric plane pairs. How
large these fractions are for a real family, energy, and tilt, and which grains of a
\emph{particular} map fall inside, is a property of the measured texture, computed in
section~\ref{sec:results}.

\subsection{Geared precision}
\label{sec:gearing}

The same low-tilt geometry that makes the window narrow makes the approach to it fine.
Viewed along the specimen normal, the tip of the beam vector moves on a circle of radius
$\sin t$ (figure~\ref{fig:cone}c), so an increment of stage azimuth moves the beam direction
by
\begin{equation}
\mathrm{d}\beta \;=\; \sin t \; \mathrm{d}\varphi .
\label{eq:gearing}
\end{equation}
Stage rotation is thus geared down by $1/\sin t$: a factor of about 8 at $t = 7\dg$, where a
full degree of stage rotation moves the beam only ${\sim}0.12\dg$. The relation is a purely
geometric consequence of the cone construction of figure~\ref{fig:cone}c and depends only on
the stage tilt, not on any detector or column geometry; equation~\eqref{eq:gearing} and its
use as the precision mechanism for rotation-stage targeting are derived here, not taken from
prior work. The two band edges of a
$\{111\}$ reflection, separated by $2\thB$ in beam angle, are consequently of order $10\dg$
apart in stage rotation ($\thB/\sin t \approx 9.7\dg$ from centre to edge at the gearing
limit). A condition selected by the criterion above can therefore be approached, and held,
with sub-$0.1\dg$ beam precision using an ordinary rotation stage. That precision
\emph{requirement}, sub-half-degree alignment for reliable defect contrast, is Qaiser et
al.'s measured figure \cite{qaiser2026}; the gearing shows it is reached without a channeling
mode, beam rocking, or a large-area detector. The gearing buys angular
\emph{resolution}, not absolute accuracy: where the beam actually sits remains limited by the
orientation map and the stage calibration feeding the computation.

\section{GRADAR: sweep, detect, rank, range}
\label{sec:method}

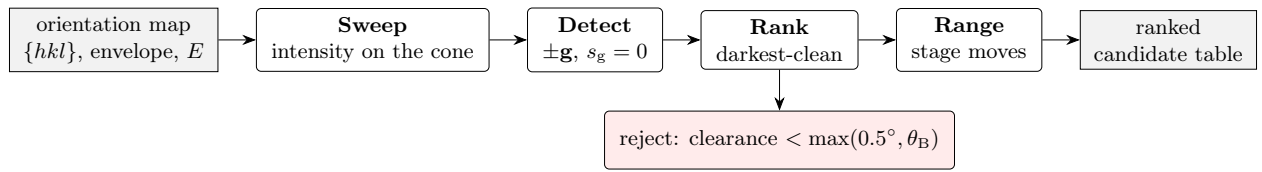
\begin{figure}[tbp]
\centering
\resizebox{\textwidth}{!}{%
\begin{tikzpicture}[
  box/.style={draw, rounded corners=2pt, align=center, minimum height=8.5mm, inner xsep=2.2mm, font=\footnotesize},
  io/.style={draw, align=center, minimum height=8.5mm, inner xsep=2mm, font=\footnotesize, fill=black!5},
  >={Stealth[length=2mm]}, node distance=4.2mm and 5.5mm]
\node[io] (in) {orientation map\\ $\{hkl\}$, envelope, $E$};
\node[box, right=of in] (sweep) {\textbf{Sweep}\\ intensity on the cone};
\node[box, right=of sweep] (detect) {\textbf{Detect}\\ $\pm\gv$, $\sg = 0$};
\node[box, right=of detect] (rank) {\textbf{Rank}\\ darkest-clean};
\node[box, right=of rank] (range) {\textbf{Range}\\ stage moves};
\node[io, right=of range] (out) {ranked\\ candidate table};
\node[box, below=6mm of rank, fill=red!8] (rej) {reject: clearance $< \max(0.5\dg,\thB)$};
\draw[->] (in) -- (sweep);
\draw[->] (sweep) -- (detect);
\draw[->] (detect) -- (rank);
\draw[->] (rank) -- (range);
\draw[->] (range) -- (out);
\draw[->] (rank) -- (rej);
\end{tikzpicture}%
}
\caption{The pipeline. Sweep simulates each grain's channeling intensity and beam direction
over the declared tilt grid and rotation window; Detect finds every crossing of the nominated
family's $\sg = 0$ locus, keeping the signed reflection; Rank orders candidates by predicted
darkness subject to the rival-clearance gate of equation~\ref{eq:clearance}, which discards
candidates outright; Range converts the survivors to dial-ready stage coordinates.}
\label{fig:pipeline}
\end{figure}

The criterion of section~\ref{sec:geometry} becomes an instrument in four steps
(figure~\ref{fig:pipeline}). The inputs are a segmented orientation map, a nominated family
$\{hkl\}$, a stage envelope (a tilt grid and rotation window, so that candidates outside what
the stage can reach are never proposed), and the beam energy.

\emph{Sweep.} For every grain, the channeling intensity is simulated over the full rotation at
each tilt of the grid, alongside the per-angle beam direction in crystal coordinates
(\texttt{rotation\_intensity\_map}). The intensity comes from a dynamically simulated
channeling reference pattern sampled at the incident beam direction, the standard forward
machinery of modern Kikuchi simulation \cite{lafond2018,zhang2026,lenthe2025}, blurred by
the measured beam convergence. This is the same forward model that eCHORD constructs; the
difference is the direction of use, taken up in section~\ref{sec:discussion}.

\emph{Detect.} Along each grain's sweep, every stage angle where a band edge of the nominated
family crosses $\sg = 0$ is located (\texttt{two\_beam\_rotations}), treating the $360\dg$
sweep as circularly closed. The signed $hkl$ is kept throughout: the $+\gv$ and $-\gv$
crossings are the band's two edges, and conflating them would discard exactly the information
a Burgers-vector analysis needs.

\emph{Rank.} The intuitive rule is to take the darkest candidate, since dark matrix means
strong channeling. It fails on its own geometry. At $\sg = 0$ every same-family candidate is
equivalent with respect to its \emph{own} band: each sits exactly $\thB$ off its plane, so the
band itself contributes identically to all of them. What differs between candidates is the
rest of the pattern, the shoulders of \emph{rival} bands passing nearby. Within-grain
intensity differences among same-family candidates therefore measure contamination, not
quality: the darker a same-family candidate, the closer it tends to sit to a rival's own
Bragg condition, a correlation quantified on a measured map in section~\ref{sec:results}.
Darkest-first selection preferentially picks
multi-beam-contaminated conditions, the opposite of what two-beam imaging wants.

\begin{figure}[tbp]
\centering
\includegraphics[width=\textwidth]{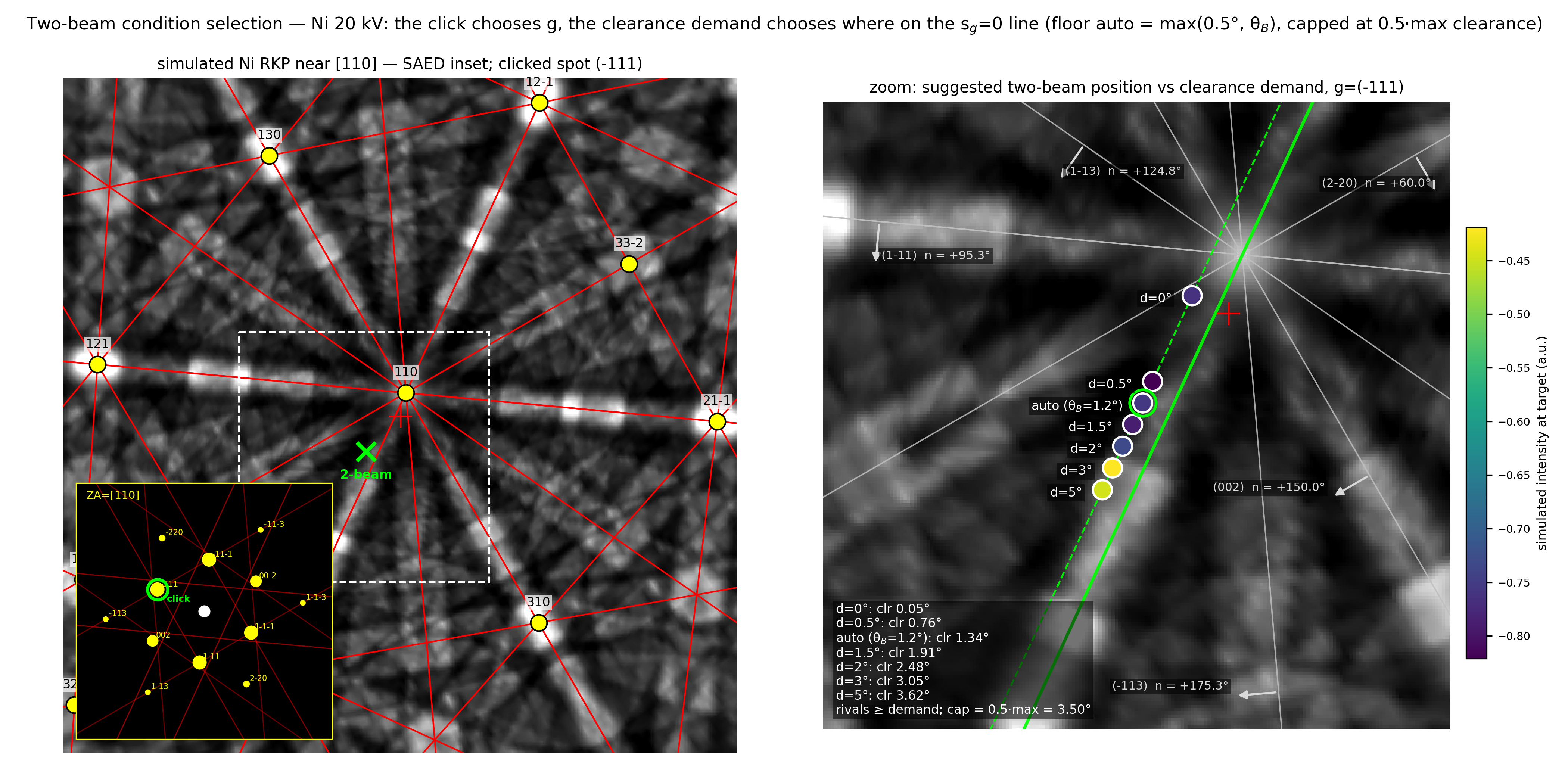}
\caption{Two-beam condition selection in \emph{autoECCI}'s interactive targeting report,
which suggests the stage tilt and rotation for the selected diffraction vector, on a
simulated nickel channeling pattern near
$[110]$ (20~kV). Left: the simulated pattern with its geometrical band overlay and
labelled zone axes; the inset kinematic diffraction net takes the click that selects the
$(\bar{1}11)$ reflection (green ring), the green cross marks the two-beam target, and the
dashed box is enlarged on the right. Right: the $(\bar{1}11)$ band centre (solid green),
its $\sg = 0$ band edge (dashed), rival bands as faint traces with plane-normal arrows,
and the suggested positions as the demanded clearance grows from $0\dg$ to $5\dg$,
coloured by simulated intensity; the green ring is the automatic-floor selection and the
text block lists the achieved clearances.}
\label{fig:selection}
\end{figure}

GRADAR therefore ranks \emph{darkest-clean} (\texttt{gvector\_targets}). Each candidate is
scored by its clearance
\begin{equation}
c \;=\; \min_{r}\,\bigl|\,\gamma_r - \theta_{\mathrm{B},r}\,\bigr| ,
\qquad
\gamma_r \;=\; \bigl|\,90\dg - \angle(\hat{\mathbf{k}}, \hat{\mathbf{g}}_r)\,\bigr| ,
\label{eq:clearance}
\end{equation}
the minimum, over every rival reflection $r$, of the deviation of the beam's glancing angle
$\gamma_r$ to the rival plane from that rival's own Bragg angle $\theta_{\mathrm{B},r}$
(kinematically forbidden rivals are dropped by structure factor). Clearance is an angular
distance in beam space, not a spatial one: $c = 0$ means some rival is itself exactly at
Bragg. Candidates with $c$ below a floor are discarded outright; the floor defaults to
$\max(0.5\dg, \thB)$, at least the excited reflection's own Bragg angle, which forces the
selection out of crowded zone-axis regions where every band crosses, and it can be set by the
user. Figure~\ref{fig:selection} shows the gate driving an
interactive selection in \emph{autoECCI}'s targeting report, which suggests the stage tilt
and rotation for a chosen diffraction vector, here on a simulated nickel channeling pattern
(kikuchipy nickel reference pattern \cite{kikuchipy}) with the beam axis $2.5\dg$ from the
$[110]$ zone axis. Interactive spot selection is inherited from the original openECCI
workflow \cite{xu2024}; the band position itself is now located deterministically. Clicking
a spot of the inset kinematic diffraction net, its spot sizes proportional to the relative
kinematic intensity $|F|^{2}$, selects the signed reflection to excite, here $(\bar{1}11)$,
and the report computes the two-beam direction from the demanded clearance, the selected
$\gv$, and the simulated intensity: every candidate position on that band's $\sg = 0$ edge,
the exact Bragg-cone locus for
$\thB = 1.21\dg$, is scored by equation~\eqref{eq:clearance}. As the demanded clearance
rises from $0\dg$ through the automatic floor at $\thB$ to $5\dg$, the returned
position walks along the edge away from the crowded zone-axis region, saturating at half
the maximin clearance, $3.50\dg$ here; the candidate colours give the simulated pattern
intensity the ranking trades. The same mechanism scores the band crossings of GRADAR's
Rank step and computes its stage targets; GRADAR itself only finds the grains that can
reach the nominated family and ranks them.

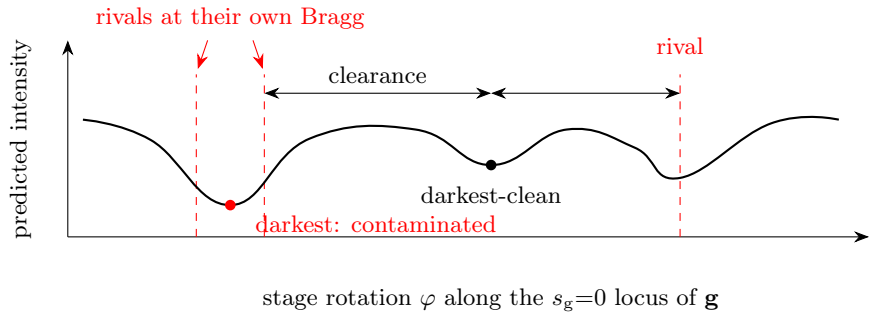
\begin{figure}[tbp]
\centering
\begin{tikzpicture}[>={Stealth[length=2mm]}, scale=1.0]
  \draw[->] (0,0) -- (10.6,0);
  \node[below] at (5.6,-0.55) {\footnotesize stage rotation $\varphi$ along the $\sg{=}0$ locus of $\gv$};
  \draw[->] (0,0) -- (0,2.6);
  \node[rotate=90, anchor=south] at (-0.35,1.3) {\footnotesize predicted intensity};
  \foreach \x in {1.7, 2.6, 8.1}{
    \draw[dashed, red] (\x,0) -- (\x,2.15);
  }
  \node[red, anchor=south] at (2.15,2.62) {\footnotesize rivals at their own Bragg};
  \draw[red,->] (1.85,2.60) -- (1.72,2.22);
  \draw[red,->] (2.45,2.60) -- (2.58,2.22);
  \node[red, anchor=south] at (8.1,2.28) {\footnotesize rival};
  \draw[thick] plot[smooth, tension=0.75] coordinates
    {(0.2,1.55) (1.1,1.28) (2.15,0.42) (3.2,1.30) (4.6,1.42) (5.6,0.95) (6.6,1.42) (7.5,1.18) (8.1,0.78) (9.3,1.5) (10.2,1.55)};
  \fill[red] (2.15,0.42) circle (0.07);
  \node[red, anchor=north west] at (2.35,0.42) {\footnotesize darkest: contaminated};
  \fill[black] (5.6,0.95) circle (0.07);
  \node[anchor=north] at (5.6,0.80) {\footnotesize darkest-clean};
  \draw[<->] (2.6,1.9) -- (5.6,1.9);
  \node[above] at (4.1,1.9) {\footnotesize clearance};
  \draw[<->] (5.6,1.9) -- (8.1,1.9);
\end{tikzpicture}
\caption{Why darkest-first fails, schematically. Along the $\sg = 0$ locus of the nominated
reflection, every candidate excites $\gv$ identically; intensity differences between
candidates come from rival bands' shoulders (dashed lines: rivals at their own Bragg
condition). The globally darkest candidate typically sits where rival shoulders pile up;
GRADAR discards candidates inside the clearance floor and takes the darkest survivor.}
\label{fig:clearance}
\end{figure}

Survivors are ordered darkest-first \emph{within} the gate, the corrected selection rule
drawn schematically in figure~\ref{fig:clearance}; section~\ref{sec:results} shows the
gate at work on a measured grain.

\emph{Range.} Each winning $(\text{grain}, \varphi, t)$ converts to dial-ready values by the
exact inverse of the stage kinematics (\texttt{pixel\_to\_stage\_angles}), for an azimuthal
rotation--tilt stage or a double-tilt holder. Two conventions matter to an operator: reported
tilts are non-negative, with direction carried by the rotation; and the dialled tilt is
conjugated by the absolute stage rotation, whose tilt axis turns with it. The full arithmetic
is part of the software's documentation rather than this paper.

\FloatBarrier

\section{Computed results on a 163-grain map}
\label{sec:results}

The computation was run on the openECCI reference dataset \cite{xu2024,openeccidata}: an
EBSD orientation map of a polycrystalline fcc austenitic stainless steel, acquired at 20~kV
and containing 163 indexed grains. The stage and detector geometry of that dataset had been
calibrated beforehand with the openECCI calibration workflow \cite{xu2024}. Each
grain's measured orientation was evaluated against equation~\eqref{eq:window} over the
declared envelope, the operative content of the Sweep and Detect steps, with the Bragg
angles of table~\ref{tab:reach}. No
imaging experiment enters this section: the inputs are the map, the energy, and the geometry,
and every value is computed.

The isotropic expectation follows from equation~\eqref{eq:window} in closed form. For
$\{111\}$ ($\thB \approx 1.17\dg$ at 20~kV) at $7\dg$ tilt, a single plane normal lands in
the annulus with probability $\sin(7\dg + 1.17\dg) = \sin(8.17\dg) = 0.142$, or 14.2\%, and
the union over the four $\{111\}$ plane pairs is bounded above by $4 \times 0.142 = 0.568$,
about 57\%. The exact isotropic value sits between the two: a Monte-Carlo evaluation over
$2\times10^{5}$ uniformly random orientations gives 48.6\% of untextured grains with a
reachable $\{111\}$ condition at this tilt, the random-texture baseline of
table~\ref{tab:reach}.

\begin{table}[tbp]
\centering
\caption{Fraction of grains with a reachable condition (equation~\ref{eq:window}), by family
and stage tilt: the measured 163-grain map versus the random-texture Monte-Carlo baseline.
Dashes: not computed.}
\label{tab:reach}
\begin{tabular}{lccccccc}
\toprule
 & & \multicolumn{3}{c}{measured map} & \multicolumn{3}{c}{random texture} \\
\cmidrule(lr){3-5}\cmidrule(lr){6-8}
Family & $\thB$ (deg) & $7\dg$ & $10\dg$ & $15\dg$ & $7\dg$ & $10\dg$ & $15\dg$ \\
\midrule
$\{111\}$ & 1.17 & 38.7\% & 52.8\% & 77.3\% & 48.6\% & 62.1\% & 79.0\% \\
$\{220\}$ & 1.91 & 79.8\% & --- & --- & 72.3\% & 86.8\% & 99.3\% \\
$\{311\}$ & 2.24 & 91.4\% & --- & --- & 92.9\% & 98.7\% & 100.0\% \\
\bottomrule
\end{tabular}
\end{table}

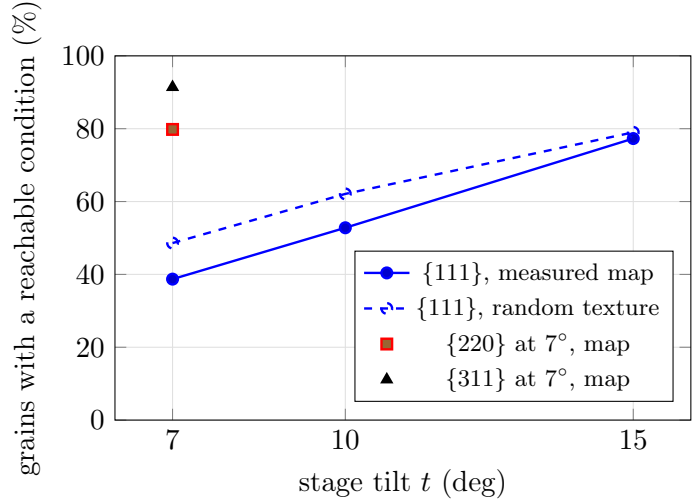
\begin{figure}[tbp]
\centering
\begin{tikzpicture}
\begin{axis}[
  width=9.2cm, height=6.4cm,
  xlabel={stage tilt $t$ (deg)}, ylabel={grains with a reachable condition (\%)},
  xmin=6, xmax=16, ymin=0, ymax=100,
  xtick={7,10,15}, grid=major, grid style={black!12},
  legend style={at={(0.97,0.05)}, anchor=south east, font=\footnotesize, row sep=1pt},
  every axis plot/.append style={line width=0.9pt}]
\addplot+[mark=*, color=blue] coordinates {(7,38.7) (10,52.8) (15,77.3)};
\addlegendentry{$\{111\}$, measured map}
\addplot+[mark=o, dashed, color=blue] coordinates {(7,48.6) (10,62.1) (15,79.0)};
\addlegendentry{$\{111\}$, random texture}
\addplot+[only marks, mark=square*, color=red] coordinates {(7,79.8)};
\addlegendentry{$\{220\}$ at $7\dg$, map}
\addplot+[only marks, mark=triangle*, color=black] coordinates {(7,91.4)};
\addlegendentry{$\{311\}$ at $7\dg$, map}
\end{axis}
\end{tikzpicture}
\caption{Reachability on the measured 163-grain austenitic steel map at 20~kV, with the
random-texture Monte-Carlo baseline for $\{111\}$ (dashed). All values are computed via
equation~\ref{eq:window}.}
\label{fig:reach}
\end{figure}

Table~\ref{tab:reach} and figure~\ref{fig:reach} give the result. Both headline consequences
would be hard to guess at the microscope. First, the $\{111\}$ fraction \emph{doubles} with a
modest tilt change, from 38.7\% of grains at $7\dg$ to 77.3\% at $15\dg$: whether a mapped
grain population is mostly usable or mostly unusable for a nominated $\{111\}$ analysis is
decided by a stage parameter chosen before the session starts. Second, family behavior
inverts the single-crystal intuition at low tilt: $\{220\}$ and $\{311\}$, with larger Bragg
angles and higher multiplicities, are reachable in 79.8\% and 91.4\% of grains at a tilt
where $\{111\}$ reaches barely a third. An operator restricted to $7\dg$, a realistic limit
for large specimens near a detector, serves the analysis better by nominating a higher-order
family than by hunting $\{111\}$ conditions that mostly do not exist.

The random-texture columns answer how far these numbers travel. This map's 38.7\% for
$\{111\}$ at $7\dg$ sits ten points below the isotropic expectation computed above, while
its $\{220\}$ value sits seven points higher than the isotropic 72.3\%.
Texture moves reachability in either direction, which is why the computation runs on the
measured map rather than on an isotropic assumption.

\begin{figure}[tbp]
\centering
\includegraphics[width=\textwidth]{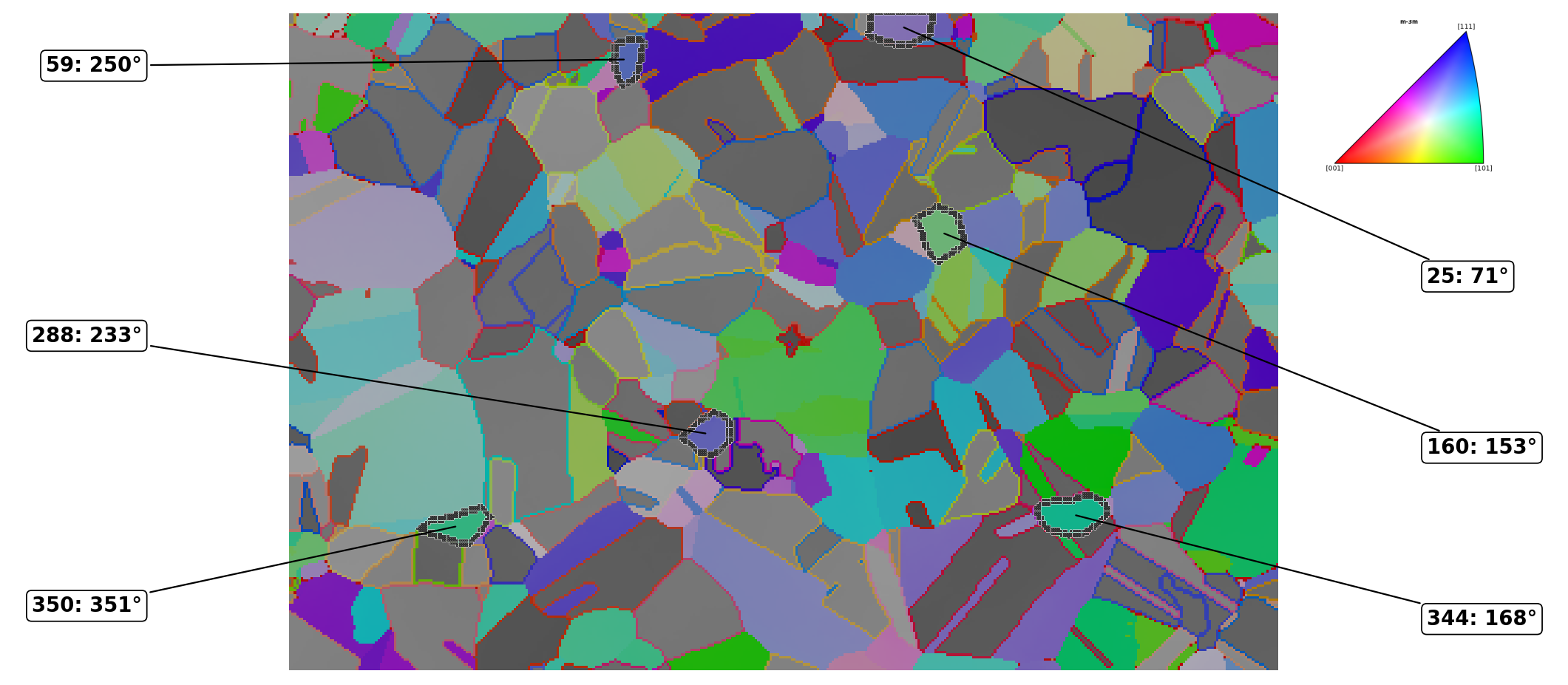}
\caption{The austenitic steel orientation map, IPF-Z coloured with a pattern-quality
overlay (the map file's \texttt{iq} column). Grains with no reachable $\{111\}$ condition
at $7\dg$ tilt are greyed out, the exclusions of equation~\ref{eq:window} drawn on the
microstructure; coloured grains are reachable. The six selected grains of
table~\ref{tab:cand} are outlined, each callout giving the grain number and the stage
rotation of its selected $\{111\}$ two-beam condition; the IPF-Z colour key is inset top
right. Generated by the reproduction notebook of appendix~\ref{app:repro}.}
\label{fig:map}
\end{figure}

The criterion also excludes, and the exclusions are falsifiable. Grains whose $\{111\}$
normals all miss the annulus at $7\dg$ are reported as unreachable purely from their measured
orientation, a prediction made blind to any image and checkable by rotating through a full
turn at that tilt and observing no $\{111\}$ band-edge crossing. Figure~\ref{fig:map} shows
these exclusions on the map itself: the greyed grains are exactly the population the
criterion rules out at this tilt.

Table~\ref{tab:cand} shows the instrument's actual output: an excerpt of the ranked
$\{111\}$ candidate table for this map at $7\dg$ tilt, in the schema the software emits, with
the six surviving candidates of largest rival clearance, one best candidate per grain. The
signed $g$ identifies which band edge is excited; the intensity $z$ is the predicted
channeling intensity standardized against the grain's own rotation curve, negative meaning
darker than that grain's mean; stage $X$ and $Y$ drive the grain's centroid under the beam.
Every value is a prediction. One feature is worth pausing on: four of the six selected
candidates are \emph{brighter} than their grain's rotation average, which is the clearance
gate working as designed. After contaminated candidates are discarded, the darkest survivor
need not be dark in absolute terms; raw darkness has been traded deliberately for a clean
condition. The table is the scientific record an acquisition session would start from.
Figure~\ref{fig:map} places the same six grains on that map: the IPF-Z-coloured
orientation map with each selected grain outlined and annotated with its grain number and
the stage rotation of its selected $\{111\}$ condition, the session plan drawn on the
microstructure it serves.

The gate of section~\ref{sec:method} is visible in this map's own numbers. Computed across
its 244 candidate conditions, the correlation between a candidate's predicted intensity and
its rival clearance is $r = +0.40$, confirming the contamination mechanism: darker
candidates sit systematically closer to rivals' own Bragg conditions. Grain 5 shows the
gate at work: of its four
$\{111\}$ candidates, the two darkest ($z = -0.64$ and $-0.29$) lie $0.24\dg$ and $1.11\dg$
from a rival's Bragg condition and are rejected by the $1.18\dg$ floor, and the selected
candidate is nearly neutral ($z = -0.004$) at $4.45\dg$ clearance.

\begin{table}[tbp]
\centering
\caption{Six highest-clearance surviving $\{111\}$ candidates at $t = 7\dg$, one best
candidate per grain, ordered by decreasing clearance. Every value is a prediction from the
measured map.}
\label{tab:cand}
\begin{tabular}{ccccccc}
\toprule
Grain & $g$ & Rotation (deg) & Intensity $z$ & Clearance (deg) & Stage $X$ (mm) & Stage $Y$ (mm) \\
\midrule
59  & $1\bar{1}\bar{1}$        & 250.3 & $-0.10$ & 7.1 & $-2.413$ & $-9.195$ \\
350 & $1\bar{1}1$              & 351.0 & $+0.52$ & 7.0 & $-2.553$ & $-9.583$ \\
160 & $111$                    & 152.9 & $+0.66$ & 7.0 & $-2.153$ & $-9.339$ \\
344 & $111$                    & 168.2 & $+1.26$ & 6.4 & $-2.044$ & $-9.573$ \\
25  & $1\bar{1}1$              & 70.9  & $+0.07$ & 6.4 & $-2.187$ & $-9.167$ \\
288 & $\bar{1}11$              & 233.1 & $+0.34$ & 6.3 & $-2.345$ & $-9.506$ \\
\bottomrule
\end{tabular}
\end{table}

The clearance floor sets the trade between contamination margin and yield, and
table~\ref{tab:floor} quantifies it on this map. At the default floor of
$\thB \approx 1.18\dg$, 68 grains keep a candidate and the median selected candidate sits at
$z = -0.17$; raising the floor to $4\dg$ keeps 55 grains and lifts the median to
$z = +0.34$; by $6\dg$ only 18 grains survive. Extra clearance is bought with grains and
with the darkness of what remains, which is why the geometric default is the floor's lower
bound rather than a recommendation of more. Where the optimum lies for measured defect
contrast is an experimental question, taken up in the deferred validation study.

\begin{table}[tbp]
\centering
\caption{Clearance-floor sweep for $\{111\}$ at $7\dg$ on the map of
table~\ref{tab:reach}. Raising the floor buys contamination margin at the cost of grains and
of the darkness of the selected candidates.}
\label{tab:floor}
\begin{tabular}{ccc}
\toprule
Clearance floor (deg) & Grains with a candidate & Median $z$ of selected candidate \\
\midrule
0.50 & 69 & $-0.40$ \\
1.18 & 68 & $-0.17$ \\
2.00 & 66 & $-0.08$ \\
4.00 & 55 & $+0.34$ \\
6.00 & 18 & $+0.52$ \\
\bottomrule
\end{tabular}
\end{table}

\section{Experimental check on published Si precession data}
\label{sec:validation}

\begin{figure}[tbp]
\centering
\includegraphics[width=\textwidth]{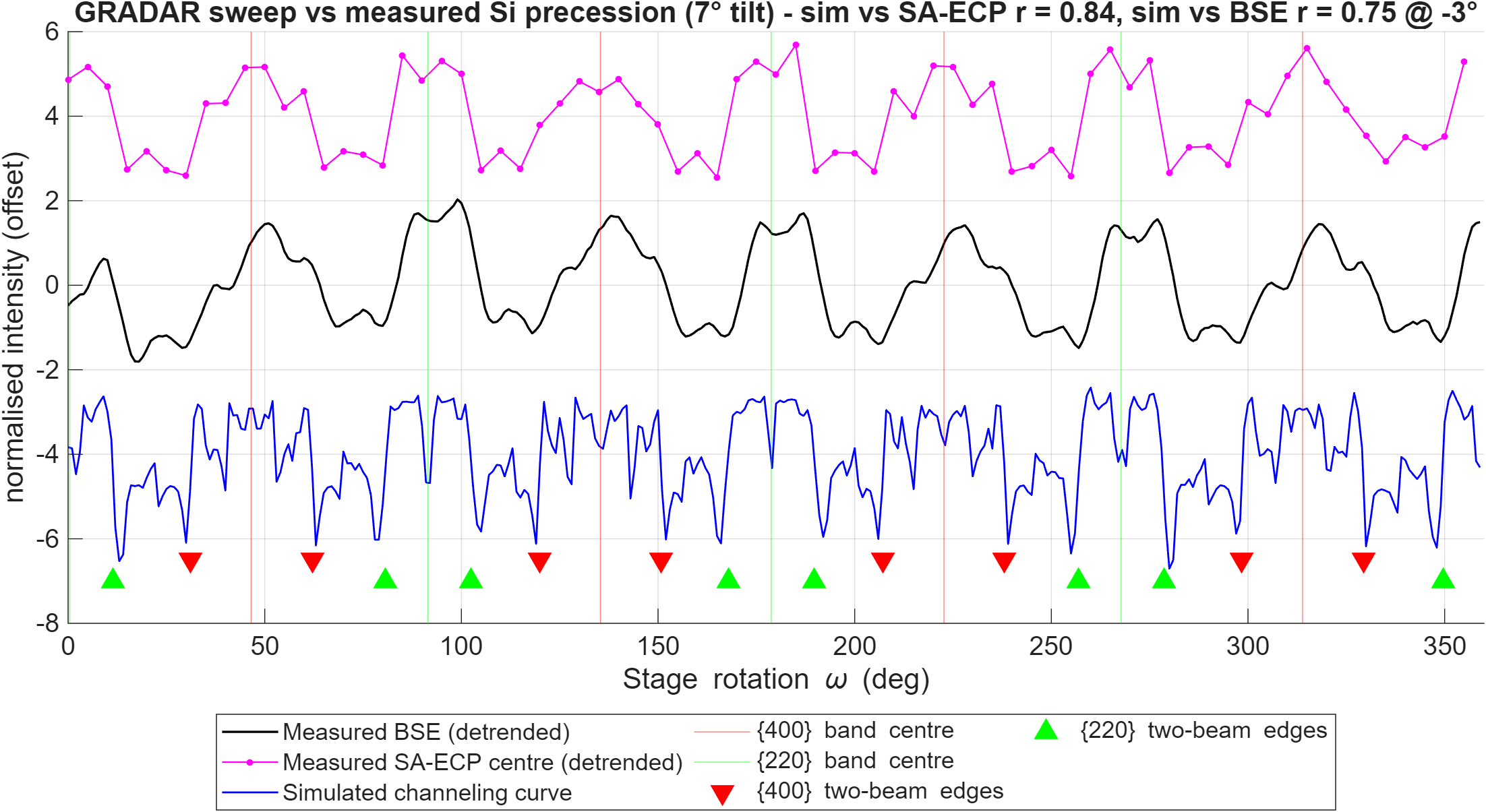}
\caption{Simulated sweep versus the measured Si precession series at ${\sim}7\dg$ tilt.
GRADAR contributes everything predicted, as the in-figure legend marks: the simulated
channeling curve (blue, the Sweep step's forward model) and the marked conditions, the
$\{400\}$/$\{220\}$ band centres (vertical lines) and two-beam edges (triangles, the
Detect step's output). The black and
magenta traces are measurements from the published dataset and are touched by no
computation beyond detrending: measured BSE trace (black), measured SA-ECP centre trace
(magenta). Simulated and measured traces agree at $r = +0.84$ (SA-ECP centre, zero lag) and
$r = +0.75$ (BSE, detrended, lag $-3\dg$).}
\label{fig:validation}
\end{figure}

The geometry of section~\ref{sec:geometry} was checked against the stage-calibration and
precession series published with the AstroECP study \cite{qaiser2026,astroecpdata}: a [001]
silicon crystal tilted by approximately $7\dg$ and rotated through a full turn on a TESCAN
AMBER-X at 20~kV, recording 73 selected-area channeling patterns in channeling mode (4Q-BSE
detector) at $5\dg$ steps together with 361 BSE micrographs in normal imaging mode at
$1\dg$ steps.
Both measured traces in figure~\ref{fig:validation} come from this published bundle. The
SA-ECP centre trace is the mean of a small central window ($8\times8$ pixels) of each
experimental pattern; the centre of an SA-ECP is the un-rocked direct beam on the optic
axis, so that central intensity is the measured channeling signal along exactly the beam
direction the sweep simulates, mirroring the central-region trace of the AstroECP study's
own precession analysis \cite{qaiser2026}. The BSE trace is the mean grey level of each
micrograph, the eCHORD-style signal \cite{lafond2018}. Only the simulated curve is computed.

The check ran on an independent implementation of the pipeline written on AstroEBSD
primitives \cite{britton2018astroebsd}, cross-checked against the Python reference
implementation below, and it is anchored to experiment at two independent levels. The first
is geometry. The 73 orientations published with the series were themselves obtained by
pattern-match indexing the same 73 experimental patterns; fitting the precession axis to
them gives a tilt of $6.86 \pm 0.06\dg$, and rotating the first frame's beam direction about
that axis with the correct stage sense reproduces all 73 indexed beam directions to at most
$0.28\dg$, while the opposite sense fails at $13.7\dg$. The cone model of
section~\ref{sec:geometry}, with the stage sense pinned empirically, is consistent with the
indexed data at the $0.3\dg$ level. The second anchor is intensity, shown in
figure~\ref{fig:validation}.

Figure~\ref{fig:validation} compares the channeling sweep GRADAR simulates from the first
frame's orientation alone, the use case of one reference orientation plus the stage model,
with the two measured traces. Against the measured SA-ECP centre intensity the simulated curve
correlates at $r = +0.84$ with no angular offset. Against the measured BSE trace, after
removal of a linear acquisition drift, the correlation is $r = +0.75$ at a lag of $-3\dg$.
All three traces share the eight-fold structure of the [001] zone. At the 16 predicted
$\{400\}$ and $\{220\}$ two-beam edges the measured BSE intensity is dark, with a mean
$z = -0.29$: precisely the property the Rank step orders on. Of the two correlations, the
SA-ECP value is the stronger validation number, because it compares simulation to
measurement at precisely the simulated beam direction, with the same detector in the same
mode. The BSE trace folds in the imaging-mode signal chain, the balance of detector gains
and the acquisition drift that had to be removed, and its residual $-3\dg$ lag is plausibly
of that instrumental origin.

The drift removal points at a general complication of BSE rotation series: brightness and
contrast do not stay put around a turn. The mean level drifts with acquisition time, and the
geometry of stage and detector adds a systematic modulation as the tilt direction sweeps
past the detector, neither of which is channeling. Here the acquisition drift was removed by
a linear detrend before comparison, and the SA-ECP centre trace, recorded by the same
detector in channeling mode, is the cross-check that survives without it. The eCHORD lineage
meets the same problem by normalizing each rotation profile before matching it into the
simulated dictionary \cite{lafond2018}, and the rotational-ECCI work of L'h\^{o}te et al.\
analysed image series across the rotation rather than single absolute levels
\cite{lhote2019}. For the deferred per-grain polycrystal validation these instrumental
intensity terms, rather than the geometry, are plausibly the limiting factor, and detector
settings that avoid clipping anywhere on the turn are a precondition for usable series.

Two implementations of the pipeline exist: the Python reference (\texttt{autoecci.gradar})
and the AstroEBSD-based port used for this check. Fed identical cone inputs for a $\{220\}$
target, the two find the same eight crossing angles within $0.0059\dg$ and agree on the
Bragg angle to $0.0005\dg$ ($1.2820\dg$ versus $1.2815\dg$, from independently maintained
physical constants), and the end-to-end run emits the candidate table in the identical
schema. The check is a single crystal: it validates the sweep, the crossing detection, and
the darkness property at predicted edges, but not the per-grain ranking across a
polycrystal, which requires the deferred validation study.

\section{Discussion}
\label{sec:discussion}

\subsection{Relation to existing tools}

The published landscape divides cleanly. The SEM lineage of TOCA/cECCI
\cite{gutierrez2013,zaefferer2014}, openECCI \cite{xu2024}, and the computed-stage route of
Mori et al.\ \cite{mori2024} solves stage kinematics for one operator-chosen grain, as does
the TEM tooling (ATEX \cite{wang2026}, ALPHABETA \cite{cautaerts2019}, $\tau$ompas
\cite{xie2020}, crystalAligner \cite{niessen2020}), whose reachable-condition screens run
within one given crystal. Patented industrial routes reach channeling information on hardware
that cannot form a selected-area pattern, by interpolating between engineered reference
regions \cite{bedell2018} or stitching composite patterns from small areas
\cite{bedell2021}, without consuming an orientation map at all. Against this record the present contribution is a
structural inversion, not a better solver: the enumeration, screening, ranking, and gating
run \emph{over grains for one nominated reflection}. Inverse stage kinematics from a known
orientation, forward Kikuchi simulation, and per-crystal reachability screens are each
solved many times over and none is claimed here.

The closest published work in intent is the orientation-adaptive virtual aperture (OAVA)
method of della Ventura et al., which applies an identical nominated diffraction condition
across a polycrystalline field of view from a single EBSD map, with no specimen tilting, by
placing virtual apertures in each pattern's reciprocal space \cite{dellaventura2025}. For
defect \emph{screening} from already-acquired pattern data, that is a strong answer to the
same motivating need, at EBSD patterns' spatial resolution and dose. It computes no stage
move and realizes no physical channeling condition: the microscope never sits at $\sg = 0$,
so the high-resolution, low-noise BSE imaging that ECCI performs there, and any subsequent
tilt series or $\gv \cdot \mathbf{b}$ workflow, remains out of its scope. GRADAR is the
complementary half: it decides where the microscope should physically go. The channeling-in
physics that both rest on is an active subject in its own right \cite{britton2026}.

The eCHORD lineage supplied this paper's forward model and its question. Lafond et al.\
construct exactly the intensity-on-a-cone machinery used here, a simulated channeling
pattern sampled along a circle of radius equal to the tilt and blurred by beam convergence,
and run it \emph{backwards}: a measured rotation profile is matched into a dictionary of a
million simulated profiles to recover orientation \cite{lafond2018,lafond2020}. Rotational
ECCI then found two-beam conditions on that geometry by eye, for one crystal, and asked in
print whether they exist for every orientation at a given tilt \cite{lhote2019}. Run
forward, tied to a measured map, the same model answers their question grain by grain;
equation~\eqref{eq:window} is that answer in closed form, and
section~\ref{sec:validation} shows the forward-run model tracking a measured precession
series on published data.

\subsection{Limitations}

The results of sections \ref{sec:geometry}--\ref{sec:results} are geometric or simulated,
and the check of section~\ref{sec:validation} is a single crystal. The reachability
percentages are properties of one measured map, material, and energy, and are not offered as
universal fractions; the random-texture baseline bounds how far they travel. The
predicted-darkness ranking is supported by the darkness of the measured intensity at the 16
predicted edges on silicon, but per-grain ranking across a polycrystal has not been
validated against measured contrast; that closure is deferred to a companion study. In
heavily deformed grains, internal lattice rotations of a degree or more mean no single
predicted condition holds across the grain, a limitation shared by every prediction-based
targeting approach (section~\ref{sec:intro}). The implementation used for the experimental
check takes its rival lists from curated reflector tables without computed structure
factors. The $\pm\gv$ assignment convention awaits experimental ground truth. The software
computes stage moves but commands no hardware. Absolute accuracy remains bounded by the
orientation map and stage calibration that feed the computation (section~\ref{sec:gearing}).

\subsection{Availability and reproducibility}

The method is implemented in \emph{autoECCI}, a Python package developed as a substantially
rewritten descendant of the open-source \emph{openECCI} project \cite{xu2024}, whose
provenance it retains under GPL-3.0-or-later; roughly three quarters of the current source is
new, under a test suite of over 960 tests. Every computed result in this paper is
reproducible by a runnable notebook shipped with the package, with the mapping given in
appendix~\ref{app:repro}. The independent implementation used for the check in
section~\ref{sec:validation}, built on AstroEBSD \cite{britton2018astroebsd}, is likewise
available. Both implementations and the notebooks are available from the author while the
public release archive of \emph{autoECCI} is being prepared. The animated version of
figure~\ref{fig:sphere} accompanies this article as supplementary movie~S1.

\section*{Acknowledgements}
The author thanks the openECCI developers, whose project this work descends from, and the
AstroECP authors for publishing the calibration and precession dataset used in
section~\ref{sec:validation}.

\section*{Declaration on the use of generative AI}

Generative AI (Anthropic Claude) was used substantially in this work, in two roles, under
continuous human direction and review.

\emph{Software.} The \emph{autoECCI} package was developed with AI assistance under a
specification-driven workflow: the author defined each capability's specification and
acceptance criteria, AI agents implemented code and tests against them, and every change
passed adversarial review, mutation testing of its guarantees, and a suite of over 960
automated tests before the author reviewed and merged it. All experimental measurements and
instrument sessions are the author's own or were taken from openly archived datasets of
published studies \cite{openeccidata,astroecpdata}, used here for validation against their
published results; the computed results of this paper are reproducible from the runnable
notebooks of appendix~\ref{app:repro}.

\emph{Manuscript.} AI assistance was used for drafting and revising the text from the
author's materials, measured results, and round-by-round written comments; all scientific
claims, data, and conclusions originate with the author, and every citation was verified
against its primary source. The author reviewed and edited all AI-assisted content and
takes full responsibility for the content of this publication.

\appendix

\section{Reproducibility of the computed results}
\label{app:repro}

Every simulation and computed value in this paper maps to a runnable Jupyter notebook in
the \emph{autoECCI} repository (\texttt{tutorials/}); the notebooks execute against the
package's public API and the named datasets, and print the paper's values beside their own
output (table~\ref{tab:repro}).

\begin{table}[htb]
\centering
\caption{Where each computed artifact is reproduced. Notebook numbers refer to
\emph{autoECCI} \texttt{tutorials/}.}
\label{tab:repro}
\resizebox{\textwidth}{!}{%
\begin{tabular}{lll}
\toprule
Artifact & Notebook & Data \\
\midrule
Candidate table generation (sweep $\to$ detect $\to$ rank $\to$ range) & 14 & openECCI map \cite{openeccidata} \\
Table~\ref{tab:reach} map columns, figure~\ref{fig:reach} & 20 & openECCI map \cite{openeccidata} \\
Table~\ref{tab:reach} random-texture baseline (MC, seed 42) & 20 & --- (computed) \\
Section~\ref{sec:results} isotropic arithmetic ($\sin(t{+}\thB)$, union bound) & 20 & --- (computed) \\
$r = +0.40$, grain-5 gate example, tables~\ref{tab:cand}--\ref{tab:floor} & 20 & tutorial-14 output \\
Figure~\ref{fig:map} (map with selected grains) & 20 & openECCI map \cite{openeccidata} \\
Section~\ref{sec:validation} sweep vs.\ measured traces & 13 & AstroECP series \cite{astroecpdata} \\
\bottomrule
\end{tabular}%
}
\end{table}

One reproducibility caveat surfaced by the notebooks is reported rather than hidden: the
grain-size filter that defines the 163-grain population runs on labels registered by an
unseeded RANSAC fit, so re-running the registration can keep a few grains more or fewer and
move the map-reachability percentages by a few points; the reproduction notebook seeds the
fit for determinism and prints the deviation from the recorded run. All statistics derived
from the shipped candidate table (the correlation, the gate example, and
tables~\ref{tab:cand}--\ref{tab:floor}) reproduce exactly.

\bibliographystyle{unsrt}
{\small
\bibliography{references}
}

\end{document}